\documentclass[conference]{IEEEtran}
\usepackage{cite}
\usepackage{amsmath,amssymb,amsfonts,mathtools,stmaryrd}
\usepackage{algorithmic}
\usepackage{bbm}
\usepackage{graphicx}
\usepackage{textcomp}
\usepackage{xcolor}
\usepackage{makecell}
\usepackage{mathrsfs}
\usepackage{tikz}
\usepackage{filecontents}
\usepackage{eurosym}
\usepackage{url}
\usepackage{array}
\usepackage[outercaption]{sidecap}
\sidecaptionvpos{figure}{t}
\usepackage{pgfplotstable} 

\DeclarePairedDelimiter\floor{\lfloor}{\rfloor}

\def\BibTeX{{\rm B\kern-.05em{\sc i\kern-.025em b}\kern-.08em
    T\kern-.1667em\lower.7ex\hbox{E}\kern-.125emX}}
\renewcommand{\IEEEQED}{\IEEEQEDopen}

\begin{document}

\title{Relay-assisted Device-to-Device Networks: Connectivity and Uberization Opportunities
}

\author{\IEEEauthorblockN{Quentin Le Gall\IEEEauthorrefmark{1}\IEEEauthorrefmark{2}, Bart\l{}omiej B\l{}aszczyszyn\IEEEauthorrefmark{2}, Elie Cali\IEEEauthorrefmark{1} and Taoufik En-Najjary\IEEEauthorrefmark{1}}
\IEEEauthorblockA{\IEEEauthorrefmark{1}Modelling and Statistical Analysis, Orange Labs Networks, Ch\^{a}tillon, France\\
Email: quentin1.legall@orange.com, elie.cali@orange.com and taoufik.ennajjary@orange.com}
\IEEEauthorblockA{\IEEEauthorrefmark{2}Inria-ENS, Paris, France
Email: quentin.le.gall@ens.fr, bartek.blaszczyszyn@ens.fr}}

\maketitle

\begin{abstract}
It has been shown that deploying device-to-device (D2D) networks in urban environments requires equipping a considerable proportion of crossroads with relays. This represents a necessary economic investment for an operator.
In this work, we tackle the problem of the economic feasibility of such relay-assisted D2D networks. First, we propose a stochastic model taking into account a positive surface for streets and crossroads, 
thus allowing for a more realistic estimation of the minimal number of needed relays.
Secondly, we introduce a cost model for the deployment of relays, allowing one to study operators' D2D deployment strategies. We investigate the example of an uberizing neo-operator willing to set up a network entirely relying on D2D and show that a return on the initial investment in relays is possible in a realistic period of time, even if the network is funded by a very low revenue per D2D user. Our results bring quantitative arguments to the discussion on possible uberization scenarios of telecommunications networks. 
\end{abstract}


\section{Introduction}
With the exponential increase of connected devices and users' intensive data demand comes the need for new ways of thinking the future cellular networks. In this regard, Device-to-Device (D2D) communication has been identified as a key component of the fifth generation (5G) of mobile networks~\cite{tehrani2014device}. In particular, by taking advantage of the densification of networks, 
D2D opens the way to cooperative networks and crowd-networking scenarios~\cite{cao2015cooperative}. \\
\indent Over the recent years, many areas of the economy have shifted towards service-based peer-to-peer business models, a process also known as uberization. In this context, the challenge of crowd-networking aided by D2D seems of critical importance for telecommunications operators.
To investigate such possibilities, mathematical models of D2D networks, more amenable to predictions and numerical simulations, have recently been introduced. \\
\indent In 5G networks, the main part of the useful spectrum consists of very high frequencies~\cite{mmWave5G} allowing only for line-of-sight (LOS) communications.  In urban environments this requires signal relaying at crossroads to ensure connectivity between adjacent streets,  a situation  often called \emph{canyon  shadowing}. In~\cite{legall2019influence}, using 
a simple stochastic percolation model, the authors have shown that large-scale connectivity requires at least 71.3\% of crossroads equipped with relays. However, the chosen  models for streets and users resulted in a zero probability of having a user located at a crossroad. In real life, one can yet expect some part of the relaying to be performed by D2D users themselves present at crossroads.

\indent In order to get a more realistic estimate of the minimal number of relays needed for connectivity, we introduce new \emph{geometric models} taking into account the surface of crossroads. Then, we introduce a related deployment and operational D2D network \emph{cost model}, allowing one to study plausible uberization scenarios of telecommunications networks through D2D. \\

\indent \emph{The main results of this paper are the following ones:}
Taking into account the possible presence of users at crossroads, we considerably reduce estimates of the minimal number of required relays. Based on our economic model, we show that a favorable return on investment is possible for an uberizing neo-operator relying on D2D only to construct its network. \\


\indent \emph{The rest of this paper is organised as follows:} We begin by recalling related works in Section~\ref{s.RelatedWorks}. Then we present our D2D connectivity  model in Section~\ref{s.Connectivitymodel}, the details of the  mathematical derivation of the geometric properties of our  crossroad models being presented  in the Appendix. The economic model for D2D costs deployment and uberization strategy is introduced in Section~\ref{S.Economicmodel}.
Finally, our results are presented in Section~\ref{s.Results} and concluding remarks are given in Section~\ref{s.Conclusions}.

\section{Related works}
\label{s.RelatedWorks}
Mathematical models of ad-hoc networks based on stochastic geometry tools go a long way back and have been introduced with Gilbert's pioneering work~\cite{gilbert1961random}. Since then, refinements of this original model taking interference into account~\cite{dousse2006percolation} have extensively been studied. The importance of the underlying geometry in mathematical models of ad-hoc networks has yet only been recently considered, based on statistical procedures proposing tools for the fitting of street systems by random tessellations~\cite{gloaguen2006fitting,courtat2011morpho}. In light of these developments, a mathematical model of D2D networks with particular attention to the modelling of geometric features and of the street system has been introduced in~\cite{cali2018percolation}. As a refinement, the impact of canyon shadowing 
has been modelled and studied in~\cite{legall2019influence,gall2019continuum}. \\
\indent Performance analyses have shown that, thanks to D2D relaying, a significant coverage extension and an enhanced quality of service (QoS) can be achieved~\cite{vanganuru2012system, GeoStoRelayD2D, fullduplexD2D, zhongD2Dcoverage, hasanrelayaided}. 
Regarding cost considerations, energy efficiency and energy costs of D2D networks are a key topic of research~\cite{xu2016energy,sun2015energy}. For instance, optimization of power allocation in underlay D2D has been studied in~\cite{yu2009power}. Mechanisms for optimisation of resource costs~\cite{ye2014resource} have also been proposed. However, all the aforementioned works do not consider the problem of connectivity and start from the assumption that connectivity of the network is already established. Moreover, to the state of our knowledge, our work is the first of its kind to investigate the use of D2D in uberization scenarios with an economic perspective. \\ 

\section{Connectivity model description}
\label{s.Connectivitymodel}
We first recall the D2D network model in a canyon shadowing urban scenario introduced in the previous work~\cite{legall2019influence}. Then, we present our new modelling of crossroads taking into account a possible population by D2D users.
\subsection{D2D network model}
\label{D2D network model}
In~\cite{legall2019influence}, a stochastic model for a D2D network in an urban scenario with the canyon shadowing assumption was introduced. The model consists of three fundamental layers:  \begin{enumerate}
    \item Streets of the city are modelled by a \emph{random tessellation}~\cite[Chapter 9]{chiu_stochastic_2013} $S$, chosen to be a Poisson-Voronoi tessellation (PVT). The average total street length per unit area, expressed in $\text{km/km}^2$, is denoted by $\gamma$. Other choices of tessellations can also be considered~\cite{gloaguen2006fitting}. 
    \item Nodes of the network consist either of D2D users equipped with mobile phones or network relays. On the one hand, D2D users are distributed on the edges of $S$ (i.e. streets of the city) according to a Cox point process~\cite[Section 5.2]{chiu_stochastic_2013} $X$ with linear intensity $\lambda$, expressed in $\text{km}^{-1}$. On the other hand, relays are placed on the vertices of $S$ (i.e. crossroads of the streets) according to a Bernoulli point process $Y$~\cite{chiu_stochastic_2013} with parameter $p \in [0;1]$. The point processes $X$ and $Y$ are also assumed to be independent conditionnally to the realisation of $S$. Contrary to previous models (e.g.~\cite{blaszczyszynd2Doptimizing}) where relays are modelled as an independent Poisson point process (PPP), the random support for streets and the canyon shadowing assumption require resorting to a Bernoulli model for relays.
    \item The D2D communication range is assumed to be a global constant $r>0$, expressed in $\text{km}$. Two nodes of the network (either D2D users or network relays) can communicate by a D2D link if and only if they are in line-of-sight (LOS), i.e. located on the same street of $S$, and the distance between them is less than $r$.
\end{enumerate}
In the aforementioned model, several assumptions have been made. First, the communication radius $r$ is considered to be a global deterministic constant, as in~\cite{glauche_continuum_2003}. This implies that the D2D users and the relays are assumed to all have the same transmission power. Moreover, note that interference phenomena and user mobility are neglected, for the sake of simplicity and mathematical tractability. In terms of relaying technique, physical and medium access technicalities (e.g. as covered by~\cite{shinrelayaidedNOMA}) are out of the scope this paper. \\
In what follows, parameters of interest will be $p$, $\lambda$ and $r$. We will consider $\gamma$ as a fixed characteristic given by geographical location. A typical value for a classical European city centre is $\gamma = 20 \, \text{km/km}^2$.\\ 
\indent The previous construction gives rise to a random connectivity graph $\mathcal{G}_{p,\lambda,r}$ which symbolises the network. Good connectivity is then mathematically interpreted as \emph{percolation} of this graph, i.e. the existence of an infinite connected component with positive probability. 

\subsection{Modelling of crossroads}
\subsubsection{Crossroad surface}
\label{sss.Crossroad}
\begin{figure}[t!]
\centering
\includegraphics[width=\linewidth]{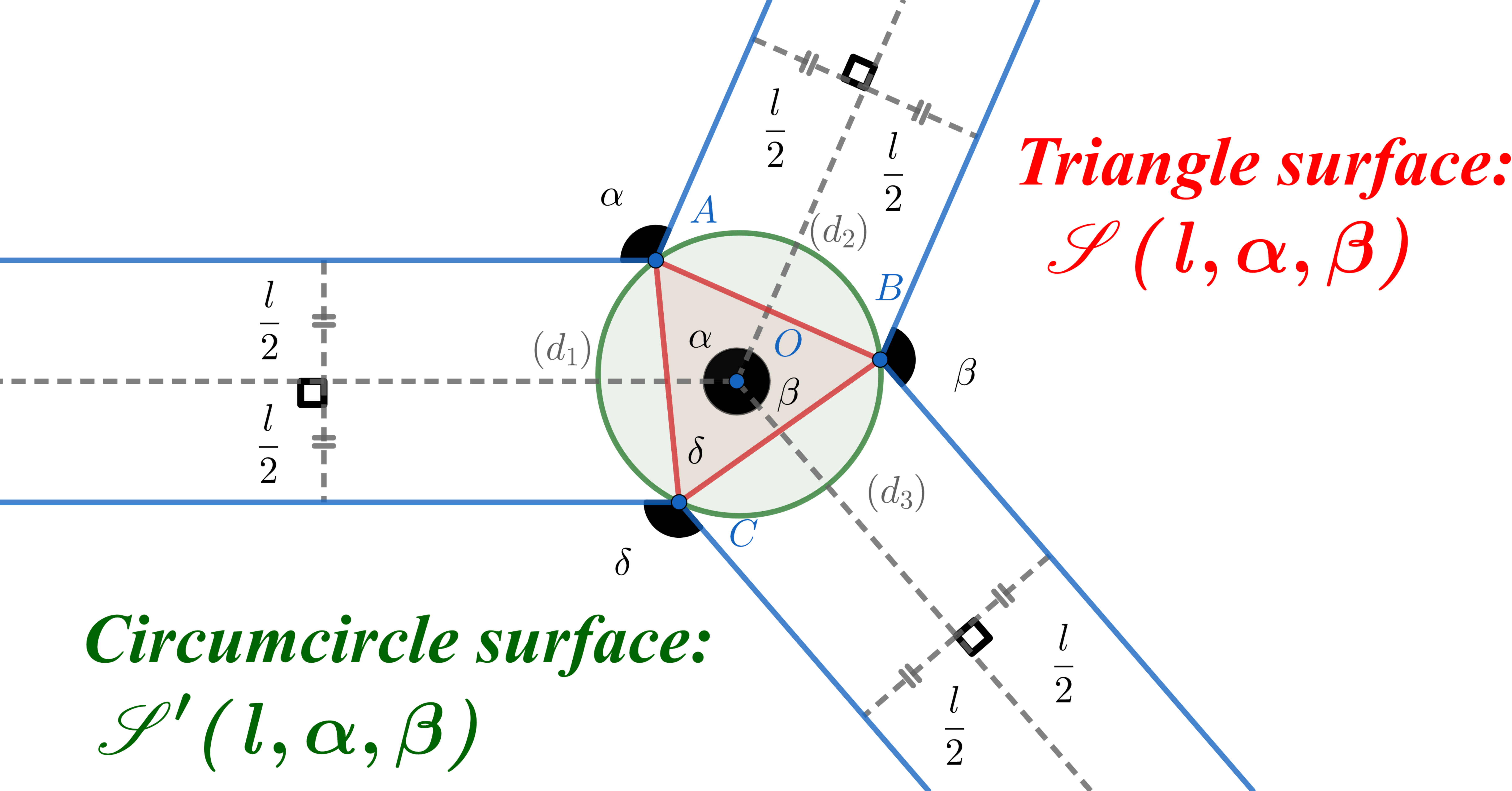}
\vspace{-1ex}
\caption{Modelling of crossroads. 
In a PVT street system, each vertex has a degree almost surely equal to 3. A crossroad is therefore defined by its location and the two random angles $\alpha \in \left[0;\pi\right], \beta \in \left[0; \pi \right] $. Dashed lines illustrate the original streets of the PVT $S$, solid blue lines represent the equivalent geometry when streets are prescribed a width $l>0$.
}
\label{fig:streets}
\vspace{-3ex}
\end{figure}

The street system $S$ being a PVT, the degree of each crossroad is almost surely equal to 3~\cite{okabe2009spatial}. Therefore, a typical crossroad,
when  centered  at the origin,
can be represented by the dashed lines $(d_1)$, $(d_2)$ and $(d_3)$ of Figure~\ref{fig:streets}, with (not necessarily equal) angles $\alpha,\beta,\delta$ being random variables.

\indent We then propose the following approach for the modelling of the typical crossroad: we prescribe a positive width $l>0$ (expressed in meters) for each street
and  imagine some geometric figure 
of non-null area (called in what follows crossroad), where a D2D user can serve as a relay between the three  adjacent streets. In order to simplify the derivation of the formula for the typical crossroad surface,  we propose two scenarios: one where the crossroad  is defined to be the triangle delimited by the intersections of the streets limits (surface $\mathscr{S}\coloneqq \mathscr{S}(l,\alpha,\beta)$ of the triangle ABC
on  Figure~\ref{fig:streets}), and one where the resulting surface is the circumcircle of this triangle (surface $\mathscr{S}'\coloneqq \mathscr{S}'(l,\alpha,\beta)$ on the same figure). Note that compared to all possible locations of relays, the first scenario corresponds to a small crossroad surface and the second to a large crossroad surface. The mathematical derivation of formulae for $\mathscr{S}(l,\alpha,\beta)$ and $\mathscr{S}'(l,\alpha,\beta)$ is presented in the Appendix. \\

\subsubsection{Typical crossroad occupation probability}
We now compute the probability that the typical crossroad is occupied, either by a physical relay or a D2D user. Denote by  $\mathcal{A} \coloneqq \mathcal{A}(l,\alpha,\beta)$ the surface of the crossroad (either $\mathscr{S}$ or $\mathscr{S}'$ in Section~\ref{sss.Crossroad}).
Using the conditional independence of users and relays given the street system 
and the Bernoulli model of the relays in the D2D network model in Section~\ref{D2D network model}, 
we pose:
\begin{align*}
\noalign{$\mathbb{P}(\text{typical crossroad is occupied})$}
&:= 1-\mathbb{P}(\text{no user in $\mathcal{A}$})\times (1-p). 
\end{align*} 
Moreover,  since the linear intensity~$\lambda$  of users in the Cox process of  
the D2D network model in Section~\ref{D2D network model} 
is  equivalent to a surface density  $\lambda / l$ in our  model where streets have width $l>0$, we pose:
\begin{equation}\label{e.no-user}  \mathbb{P}(\text{no user in $\mathcal{A}$}):=\mathbb{E}\left[e^{-\frac{\lambda}{l}\mathcal{A}(l,\alpha,\beta)}\right],
\end{equation}
where $\alpha$ and $\beta$ are random variables and denote the angles spanned by the streets of the typical crossroad of the PVT $S$. To compute the expectation in~\eqref{e.no-user}, we use the joint probability density for the random vector $(\alpha,\beta)$ (note $\delta=2\pi-(\alpha+\beta)$) 
 computed by Muche~\cite{muche1998poisson}:
\begin{gather}
 \notag  f(\alpha,\beta)=-\frac{8}{3\pi}\sin \alpha \sin \beta \sin(\alpha+\beta), \\
    \text{with } 0<\alpha<\pi,\, \pi-\alpha<\beta<\pi. \label{density-angles-typical-crossroad}
\end{gather}
Therefore, the crossroad occupation probability is given by:
\begin{gather}
    \notag F(\lambda,p,l):=\mathbb{P}(\text{typical crossroad is occupied}) \\
        \label{crossroad-occupation-probability} =1-(1-p)\int_{\alpha=0}^{\pi}\int_{\beta=\pi-\alpha}^{\pi} e^{-\frac{\lambda}{l}\mathcal{A}(l,\alpha,\beta)}f(\alpha,\beta)d\beta d\alpha,
\end{gather}
where $(\alpha,\beta)\mapsto f(\alpha,\beta)$ is the joint density given by~\eqref{density-angles-typical-crossroad} and $\mathcal{A}(l,\alpha,\beta)$ is the resulting crossroad surface, either $\mathscr{S}(l,\alpha,\beta)$ or $\mathscr{S}'(l,\alpha,\beta)$ according to the chosen modelling.

\begin{figure}[t!]
    \centering
    \includegraphics[width=0.8\linewidth]{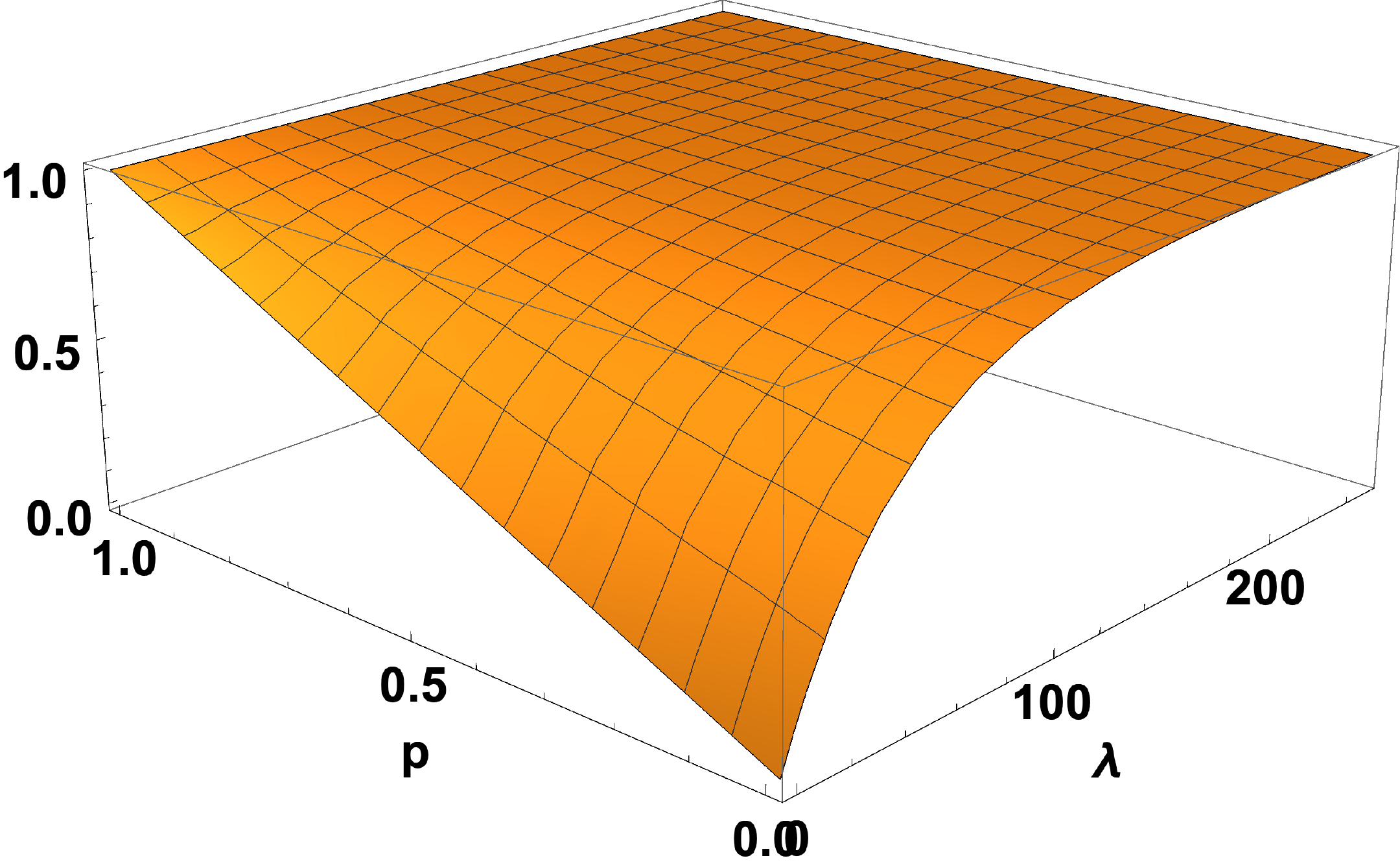}
    \vspace{-1ex}
    \caption{Crossroad occupation probability $(\lambda,p) \mapsto F(\lambda,p,l=20 \, \text{m})$ when $\mathcal{A}(l,\alpha,\beta)=\mathscr{S'}(l,\alpha,\beta)$ and $l=20 \, \text{m}$}
    \label{fig:crossroad-occupation-circle}
\vspace{-3ex}    
\end{figure}

Figure~\ref{fig:crossroad-occupation-circle} shows the plotting of the occupation probability of the typical crossroad~$F(\lambda,p,l)$ as a function of the user density~$\lambda$ and the physical relay proportion~$p$, in an example when the crossroad resulting surface is the circumcircle area and the street width $l$ is equal to 20 meters, a typical value for a classical European city centre, estimated via statistical methods suggested in~\cite{courtat2011morpho}.

\subsection{Minimal relay proportion}
\indent Using our new model taking into account the possible population of crossroads by D2D users or physical relays, 
we define the {\em minimal proportion  of physical  relays} needed to ensure large-scale connectivity of the D2D network as follows: 
\begin{equation*}
    p_c(\lambda,r) \coloneqq \inf \lbrace p \in \left[0,1\right], \, \mathcal{G}_{F(\lambda,p,l),\lambda,r}  \, \text{percolates}  \rbrace
\end{equation*}
where, recall, percolation means    existence  of  an  infinite  connected component with positive probability.
This quantity $p_c(\lambda,r)$ represents the necessary investment in relays for operators, given the D2D user density and the D2D technology. $p_c(\lambda,r)$ can be calculated using the estimation  of the  {\em minimal  proportion of occupied crossroads}   \begin{equation*}
    p^*(\lambda,r) \coloneqq \inf \lbrace p \in \left[0,1\right], \, \mathcal{G}_{p,\lambda,r} \, \text{percolates} \rbrace
\end{equation*}
originally defined and estimated in~\cite{legall2019influence} by taking $p_c(\lambda,r)=\min(1,\max(0,p))$, where $p$ is the (unique)  
solution of the equation
$$
F(\lambda,p,l)=p^*(\lambda,r).
$$

\section{Economic model description}
We now introduce a cost model for the deployment of relays in a relay-assisted D2D network. This in turn allows us to introduce the uberization strategy of a neo-operator willing to set up a network by entirely relying on D2D technology.
\label{S.Economicmodel}
\subsection{Cost model}
\label{Ss.Costmodel}
Relevant parameters of our cost model for the deployment of a relay-assisted D2D network are presented in Table~\ref{tab:cost-parameters}. In particular, regarding the costs of relays, two types of costs are to be paid by an operator:
 capital expenditures (CAPEX) per relay ($c_{\text{CAPEX}}$) and  operational expenditures (OPEX) per relay, per year  ($c_{\text{OPEX}}$) (covering maintenance, energy, etc.), to be taken into account during network's exploitation.
Due to the fact that relays depreciate over time, we consider a relay depreciation period $T_{\text{DEP}}$ of years, after which relays will successively be replaced. 



Regarding the operator's revenue on the use of its D2D network, several sources  are possible. For instance, the use of the D2D service can be billed to users as a monthly subscription fee. Another source of revenue for the operator can consist in funding the deployment of its D2D network by advertisement. For simplicity, we do not take into account the nature of the revenue for the operator and model it by a parameter $G \geq 0$ corresponding to the monthly revenue received by the operator per D2D user.
\\
\begin{table}[t!]
\caption{Relevant cost parameters and their signification}
\begin{center}
\begin{tabular}{|l|>{\centering\arraybackslash}m{6cm}|}
\hline
\thead{\textbf{Parameter}}        &  \thead{\textbf{Description}}   \\ \hline
\makecell{$c_{\text{CAPEX}}$} & \makecell{CAPEX cost of one physical relay}\\ \hline
\makecell{$c_{\text{OPEX}}$} & \makecell{Yearly OPEX cost of one physical relay} \\ \hline
\makecell{$\eta$} & \makecell{$c_{\text{OPEX}}/c_{\text{CAPEX}}$} \\ \hline
\makecell{$G$} & \makecell{Operator's monthly revenue per D2D user} \\ \hline
\makecell{$T_{\text{DEP}}$} & Depreciation period of one physical relay (in years) \\ \hline
\makecell{$\mathscr{A}$} & \makecell{Area covered by the D2D network} \\ \hline
\end{tabular}
\label{tab:cost-parameters}
\end{center}
\vspace{-3ex}
\end{table}
 \indent  Moreover, we need to specify the number of relays and D2D users in the network.
 In this regard, we assume that an
area $\mathscr{A}$ of km${}^2$
 is covered by the considered D2D network. This, together
 with the previously considered network parameters
 ($\gamma$, $p$, $\lambda$)  allows us to express the 
 mean number of relays in the network 
 as $p\gamma^2\mathscr{A}/2$
 (recall from~\cite{okabe2009spatial} that $\gamma^2/2$ is the average number of vertices per unit area in the PVT)
 and the mean number of users as
 $\lambda \gamma \mathscr{A}$.

\subsection{Neo-operator's uberizing strategy}
\label{Ss.uberizing-strategy}
We now consider an operator willing to deploy its network by relying on D2D technology only (and thus limiting infrastructure investments in base stations) with the following business strategy.

In order to spread CAPEX invesments over time and not massively invest too quickly, deploying relays will be done in two steps: a first network deployment phase before the commercial launch of the service and a second one after this commercial launch, when the neo-operator starts to gain customers. 

\indent Initially, an investment in relays (CAPEX) has to be made before the commercial launch of the service. 
During this initial period of network deployment, no customers have subscribed to the service and so the user density is null ($\lambda(t)=0$). Note that once a  relay has been installed, operational costs ($c_{\text{OPEX}}$)  are to be paid for this relay by the operator. We denote by $p_{\text{min}}$ the proportion of crossroads that the neo-operator wishes to equip with relays at the end of this first phase. The choice of a value for $p_{\text{min}}$ is done by the neo-operator and can be oriented by the results about the minimal relay proportion needed for large-scale connectivity of the network. We assume that the relays are deployed in equal amount during each month of the initial deployment period, as highlighted by Table~\ref{tab:cost-relevant}. \\ 
\indent At month $t=T_{\text{LAUNCH}}$, the service of the neo-operator is commercially launched. From that moment, the operator gets a revenue $G$ per month for each D2D user.
Worth noticing is that the neo-operator might not have sufficiently many customers yet (and so D2D relays) to ensure a good QoS. Over time and by contagion effect, new customers will be attracted and will allow to reach the critical mass needed to ensure large-scale connectivity of the network.
 This will be taken into account by allowing for the user density  $\lambda(t)$ to depend on time. \\
 \indent When the D2D service is commercially launched, the neo-operator will start its second relay deployment phase, until reaching a proportion $p_{\text{max}}$ at time $t=T_{\text{CRITICAL}}$. Again, we assume that the relays are deployed in equal amount during each month of this second phase, see Table~\ref{tab:cost-relevant}.  In practice, the choice of values for $p_{\text{max}}$ and $T_{\text{CRITICAL}}$ is also oriented by the results about the minimal relay proportion needed for large-scale connectivity of the network, and done by the neo-operator in such a way that when all relays have been deployed (i.e. in proportion $p_{\text{max}}$ at time $T_{\text{CRITICAL}}$), the critical mass of users ensuring large-scale connectivity of the network has almost been reached. In other words, this means that the neo-operator tunes $p_{\text{max}}$ and $T_{\text{CRITICAL}}$ so as to have:
 \begin{equation}
 \label{tuning-parameters}
     p_{\text{max}} \approx p^*(\lambda(T_{\text{CRITICAL}}),r).
 \end{equation}
 \indent Finally, once all relays have been deployed, the neo-operator starts to replace them at time $t=T_{\text{DEP}}$,  in such a way that the whole relay fleet will entirely be replaced within another depreciation period $\left[T_{\text{DEP}};2T_{\text{DEP}}\right]$, and so on. 
\begin{table}[t!]
\caption{Relevant quantities for costs computations}
\begin{center}
\begin{tabular}{|l|>{\centering\arraybackslash}m{4.3cm}|}
\hline
\thead{\textbf{Parameter}}        &  \thead{\textbf{Description}}   \\ \hline
\makecell{$t$} & Time (in months)\\ \hline
\makecell{$\lambda(t)$} & User density at month $t$ \\ \hline
\makecell{$T_{\text{LAUNCH}}$} & Time at which the service is commercially launched\\ \hline
\makecell{$T_{\text{CRITICAL}}$} & Time at which all relays have been deployed \\ \hline
\makecell{$\floor*{\frac{p_{\text{min}}\gamma^2 \mathscr{A}}{2T_{\text{LAUNCH}}}}c_{\text{CAPEX}}$} & Monthly CAPEX spent before the launch of the commercial service \\ \hline
\makecell{$\floor*{\frac{(p_{\text{max}}-p_{\text{min}})\gamma^2 \mathscr{A}}{2(T_{\text{CRITICAL}}-T_{\text{LAUNCH}})}}c_{\text{CAPEX}}$} & Monthly CAPEX spent in the second phase of relay deployment \\ \hline
\makecell{$\frac{p_{\text{max}}\gamma^2 \mathscr{A}}{2}\frac{\eta c_{\text{CAPEX}}}{12}$} & Monthly OPEX spent once all relays have been deployed \\ \hline
\makecell{$G \lambda(t) \gamma \mathscr{A}$} & Revenue of the operator at month $t$ \\ \hline
\end{tabular}
\label{tab:cost-relevant}
\end{center}
\vspace{-3ex}
\end{table}

 In the end, the neo-operator wishes to have a return on investment (ROI) by counting on the revenue coming from its customers and a limited investment in renewing physical relays.  
 More specifically,  a first quantity of interest for the operator's strategical decisions is the monthly \emph{cash flow} $CF(t)$, which corresponds to the difference between the money earned and the money spent each month:
 \begin{equation}
     \label{Cash-Flow}
     CF(t) \coloneqq G\lambda(t)\gamma\mathscr{A}- N_{\text{B}}(t)c_{CAPEX} - N(t) \frac{\eta c_{\text{CAPEX}}}{12},
 \end{equation}
 where $N_{\text{B}}(t)$ and $N(t)$ respectively denote the number of relays that need to be bought at the beginning of month $t$ and the total number of relays in the network at the end of month $t$. 
 A second quantity of interest is the \emph{cumulated revenue} $CR(t)$ up to month $t$: 
  \begin{equation}
 \label{Cumulated-Revenue} 
 CR(t) \coloneqq \sum_{s=0}^{t} CF(s),
  \end{equation}
where $CF(s)$ is the cash flow at month $s$ defined by~\eqref{Cash-Flow}.  The ROI is reached when the cumulated revenue becomes positive. 

\section{Results}
\label{s.Results}
In this section, we present the results of our numerical estimations. We first investigate the physical relay proportion needed for large-scale connectivity of the D2D network, taking into account the presence of mobile users at crossroads. These estimates, combined with our cost model assumptions, allow us to investigate an uberizing scenario described in Section~\ref{Ss.uberizing-strategy}.


\subsection{Minimal relay proportion needed for large-scale connectivity of a D2D network}
\label{Ss.detailed.connectivity}

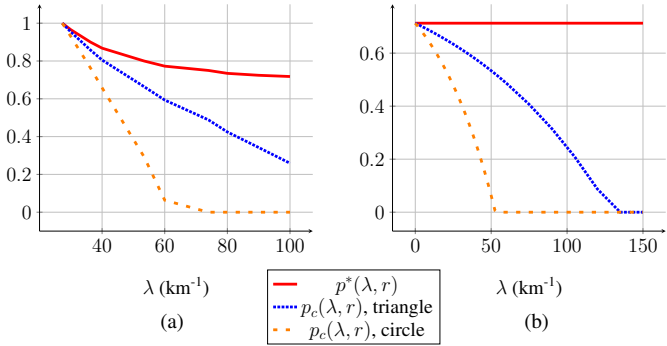
\begin{figure}[t!]
\centering
\hspace{-.5cm}
\begin{tikzpicture}[scale=0.53]
\begin{axis}[
name=ax1,
legend style={font=\Large,at={(1.15,-0.17)},anchor=north},
axis lines = left,
    ticklabel style = {font=\Large},
    xlabel= $\lambda$ (km\textsuperscript{-1}),
    xlabel style = {font=\Large, below=5mm},
    ylabel = {},
    ylabel style = {font=\Large, above=5mm},
    enlargelimits=true,
    grid=both,
]
\addlegendentry{$p^*(\lambda,r)$}
\addplot[color=red, mark= \empty, line width = 2pt] table [x=lambda,y=ptotal, col sep = semicolon, /pgf/number format/read comma as period] {pc50m.csv};
\addlegendentry{$p_c(\lambda,r)$, triangle }
\addplot[color=blue, mark= \empty, line width = 2pt, densely dotted] table [x=lambda,y=ptriangle, col sep = semicolon, /pgf/number format/read comma as period] {pc50m.csv};
\addlegendentry{$p_c(\lambda,r)$, circle}
\addplot[color=orange, mark= \empty, line width = 2pt, loosely dashed] table [x=lambda,y=pcercle, col sep = semicolon, /pgf/number format/read comma as period] {pc50m.csv};

\end{axis}

\begin{axis}[
at={(ax1.south east)},
xshift=2cm,
axis lines = left,
    ticklabel style = {font=\Large},
    xlabel= $\lambda$ (km\textsuperscript{-1}),
    xlabel style = {font=\Large, below=5mm},
    ylabel = {}, 
    ylabel style = {font=\Large, above=5mm},
    enlargelimits=true,
    grid=both,
]
\addlegendentry{$p^*$}
\addplot[color=red, mark= \empty, line width = 2pt] table [x=lambda,y=ptotal, col sep = semicolon, /pgf/number format/read comma as period] {pc200m.csv};
\addlegendentry{$p_c(\lambda,r)$, triangle case}
\addplot[color=blue, mark= \empty, line width = 2pt, densely dotted] table [x=lambda,y=ptriangle, col sep = semicolon, /pgf/number format/read comma as period] {pc200m.csv};
\addlegendentry{$p_c(\lambda,r)$, circle case}
\addplot[color=orange, mark= \empty, line width = 2pt, loosely dashed] table [x=lambda,y=pcercle, col sep = semicolon, /pgf/number format/read comma as period] {pc200m.csv};

\legend{}

\end{axis}
\end{tikzpicture}
\\[-4ex]
\centerline{\footnotesize\hspace{0.22\linewidth} (a)\hspace{0.5\linewidth} (b)\hspace{0.2\linewidth}\ }
\vspace{0ex}
\caption{Estimation of the minimal relay proportion needed for large scale connectivity as a function of user linear intensity $\lambda$. The plain red curve is the total proportion $p^*(\lambda,r)$ of equipped crossroads required for large-scale connectivity of the network. The two other curves are the proportion of physical relays $p_c(\lambda,r)$ needed, respectively in the triangle case (dotted blue) and in the circle case (dashed orange).  Left: $r = 50 \, \text{m}$ (D2D technology corresponding to WiFi). Right: $r = 200 \, \text{m}$ (D2D technology corresponding to millimeter-Wave frequencies).
}
\label{fig:real-relay-proportion}
\vspace{-3ex}
\end{figure}


\indent Figure~\ref{fig:real-relay-proportion} illustrates two examples of the estimations for this minimal physical relay proportion~$p_c(\lambda,r)$ required for connectivity of the D2D network at large scale. 
In both examples, two facts are of noticeable importance. On the one hand, taking into account the presence of D2D users being able to act as relays on crossroads considerably reduces the former estimates ($p^*(\lambda,r)$) provided in~\cite{legall2019influence}. 
On the other hand, the influence of the chosen geometry for the modelling of crossroads is crucial: the estimated physical relay proportion required for large-scale connectivity varies by at least a factor two in high user density scenarios. This is due to the fact that the ratio between the area of a triangle and its circumcircle is considerably small. 
As a matter of fact, it is much more likely to find mobile users at crossroads in the circumcircle case than in the triangle case, which results in less physical relays needed at a global scale. \\
\indent Figure~\ref{fig:real-relay-proportion}(a) presents the estimates when the D2D range $r=50 \, \text{m}$, which corresponds to a scenario where the radio technology supporting D2D would be WiFi~\cite{asadi2014modeling}. In such a case, note that a sufficiently high density of users, mainly $\lambda \geq 60 \, \text{km}^{-1}$, can fully compensate the relays in the circumcircle case. Note that such a threshold is not unrealistic, as $\lambda \approx 60 \, \text{km}^{-1}$ equivalently means that there are in average 60 D2D users per kilometer of streets, which means less than one user every fifteen meters in average. 

The case of larger connectivity radii is for instance depicted in Figure~\ref{fig:real-relay-proportion}(b), where $r=200 \, \text{m}$.
From the previous work~\cite{legall2019influence}, we know  that for such  regimes, we approach an asymptotic situation where the total proportion of equipped crossroads $p^*(\lambda,r)$ is independent of $\lambda$ and $r$ and equal to the absolute minimal proportion of $71.3\%$ of crossroads
(cf. the horizontal red line on Figure~\ref{fig:real-relay-proportion}(b)).
Such high values of $r$ require the use of  millimeter-Wave frequencies~\cite{mmWave5G} for D2D communications. This will be the case in 5G networks, where one will thus approach the asymptotic situation depicted in Figure~\ref{fig:real-relay-proportion}(b).


\subsection{Analysis of an uberizing neo-operator's strategy}
\label{Ss.uberizing.operator}
\begin{table}[t!]
\caption{Simulation parameters}
\begin{center}
\begin{tabular}{|l|>{\centering\arraybackslash}m{5.2cm}|} \hline
\thead{\textbf{Simulation parameter}}        &  \thead{\textbf{Numerical value}}   \\ \hline
\makecell{$c_{\text{CAPEX}}$} & \makecell{\euro 1200}\\ \hline
\makecell{$\eta$} & \makecell{10\%} \\ \hline
\makecell{$G$} & \makecell{\euro 3} \\ \hline
\makecell{$T_{\text{DEP}}$} & \makecell{7 years = 84 months } \\ \hline
\makecell{$T_{\text{LAUNCH}}$} & \makecell{1 year = 12 months} \\ \hline\makecell{$T_{\text{CRITICAL}}$} & \makecell{2 years 1/2 = 30 months} \\ \hline
\makecell{$p_{\text{min}}$} & \makecell{10\%} \\ \hline
\makecell{$p_{\text{max}}$} & \makecell{20\%} \\ \hline
\makecell{$\lambda(t)$} & \makecell{As in Fig.~\ref{fig:user-growth-density}} \\ \hline
\makecell{$\gamma$} & \makecell{20 km/km\textsuperscript{2}} \\ \hline
\makecell{$\mathscr{A}$} & \makecell{25 km\textsuperscript{2}} \\ \hline
\end{tabular}
\label{tab:cost-quantities-parameters}
\end{center}
\vspace{-3ex}
\end{table}

\begin{figure}[t!]
\centering
    \begin{tikzpicture}[scale=0.55]
\begin{axis}[
    axis lines = left,
    xticklabel style = {rotate = 45},
    ticklabel style = {font=\Large},
    xlabel= $t$ (months),
    xlabel style = {font=\Large, below=5mm},
    ylabel = {$\lambda(t)$ (km\textsuperscript{-1})},
    ylabel style = {font=\Large, above=5mm},
    enlargelimits=true,
    grid=both,
]
\addplot [
domain=0:12,
samples=100,
color=blue,
line width = 2pt,
]
{0};

\addplot [
domain=12:120,
samples=100,
color=blue,
line width = 2pt,
]
{0.2*((90/(0.5*(1+200*exp(-0.5*(x-11.961)))))+100*(1-exp(-(x-11.961)/36)))};

\end{axis}
\end{tikzpicture}    
    \caption{User density as a function of time. }
    \label{fig:user-growth-density}
    \vspace{-3ex}
\end{figure}
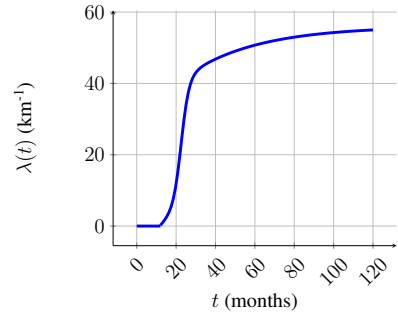

\indent We performed a numerical evaluation of the uberizing strategy presented in Section~\ref{Ss.uberizing-strategy} with parameters coming from internal data provided in Table~\ref{tab:cost-quantities-parameters} and a user density $\lambda(t)$ illustrated by Figure~\ref{fig:user-growth-density}. We chose a user density $t \mapsto \lambda(t)$ which is increasing as a function of time, thus assuming that the neo-operator does not lose customers. Note the user density is null before commercial launch, i.e. for $t\leq T_{\text{LAUNCH}}=1 \, \text{year}$, then undergoes a steep increase for a six-month period corresponding to early adopters and finally undergoes a slower growth once a critical mass of D2D users has been attained. 


In our numerical example, we took $p_{\text{min}}=10\%$ and  $p_{\text{max}}=20\%$.   Moreover, $T_{\text{CRITICAL}}=30 \, \text{months}$ has been chosen. Note that this choice of parameters satisfies~\eqref{tuning-parameters}. Indeed, $\lambda(T_{\text{CRITICAL}}) \approx 45 \, \text{km}\textsuperscript{-1}$. 
In a 5G millimeter-Wave context where $r=200 \, \text{m}$ (see Figure~\ref{fig:real-relay-proportion}(b)), assuming circle crossroads, this density of users requires  a minimal relay proportion of about $p_c(\lambda = 45 \, \text{km}\textsuperscript{-1}, r=200 \, \text{m}) \approx 20\%$. Thus, at the critical time $T_{\text{CRITICAL}}$, the neo-operator has fully deployed its relays and ensured large-scale connectivity of its D2D network.

\begin{figure}[t!]
\centering
\hspace{-0.8cm}
\begin{tikzpicture}[scale=0.515] 
\begin{axis}[
axis lines = left,
xticklabel style = {rotate = 45},
    ticklabel style = {font=\Large},
    xlabel= $t$ (months),
    xlabel style = {font=\Large, below=5mm},
    ylabel = {$CR(t)$ (k\euro)},
    ylabel style = {font=\Large, above=5mm},
    enlargelimits=true,
    grid=both,
]
\addplot[color=orange, mark=\empty, line width = 2pt] table [x=Time,y=Cumulated, col sep = semicolon,  /pgf/number format/read comma as period] {techeco.csv};
\end{axis}
\end{tikzpicture}
\hspace{-.2cm}
    \begin{tikzpicture}[scale=0.515] 
\begin{axis}[
axis lines = left,
xticklabel style = {rotate = 45},
    ticklabel style = {font=\Large},
    xlabel= $t$ (months),
    xlabel style = {font=\Large, below=5mm},
    ylabel = {$CF(t)$ (k\euro)},
    ylabel style = {font=\Large, above=2.5mm},
    enlargelimits=true,
    grid=both,
]
\addplot[color=red, mark=\empty, line width = 2pt] table [x=Time,y=CashFlow, col sep = semicolon,  /pgf/number format/read comma as period] {techeco.csv};
\end{axis}
\end{tikzpicture}
\vspace{-1ex}
\centerline{\footnotesize\hspace{0.3\linewidth} (a)\hspace{0.4\linewidth} (b)\hspace{0.2\linewidth}\ }
\vspace{-3ex}

\caption{Left: Cumulated revenue as a function of time. The ROI is reached at $t=43$ months. Right: Cash flow as a function of time.}
\label{fig:cumulated-revenue-cash-flow}
\vspace{-3ex}
\end{figure}
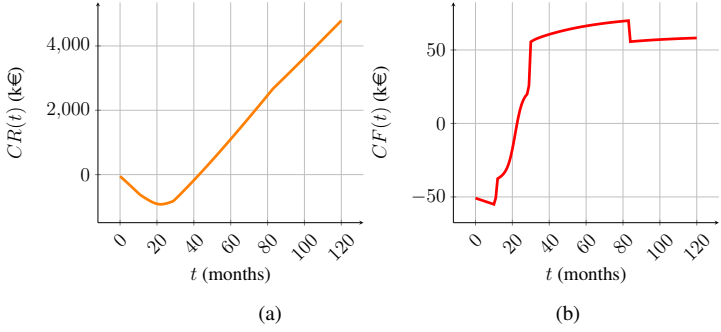

\indent Figure~\ref{fig:cumulated-revenue-cash-flow} shows the evolution of the cumulated revenue and of the cash flow over time. With the parameters values prescribed by Table~\ref{tab:cost-quantities-parameters}, we obtain a ROI of 43 months, i.e. 31 months after the commercial launch of the D2D service. Distinguishable times are the ones of the commercial launch 
(where the cash flow $CF$ starts to increase abruptly), the time of the critical connectivity 
(where the cash flow $CF$ increases abruptly for the second time)  and the time 
at which relays start to be replaced (this is when the cash flow $CF$ drops suddenly). 

\indent In practice, note that an ROI of 43 months is more than satisfying for potential investors. Moreover, our cost model is sufficiently generic so that any operator willing to launch a D2D service can replace our simulation parameters by its own internal data (e.g. obtained via marketing analyses) and hence get an idea of when setting up a relay-assisted D2D network will be profitable. This in turn can help to make strategical decisions regarding D2D investments. \\
\indent  Our cost model paves the way to interesting discussions regarding deployment of relay-assisted D2D networks. In the case of a neo-operator, note that the number of D2D users may be too small at the beginning of the service to get a fully connected network. The neo-operator would then start by offering proximity services~\cite{lin2014overview}, until large-scale connectivity of its network is ensured by sufficiently many D2D users. Another possibility to reach this critical mass of users could consist in resorting to customers of extisting operators through negociated costs.  
Finally, funding the D2D service from its very beginning is essential, otherwise the ROI would happen much later in time, and thus the uberization scenario described in our model might not be profitable for a new actor in the telecommunications sector.

\section{Concluding remarks}
Proposing more realistic geometric  models for  crossroads in urban environments, we considerably reduced predictions of the minimal proportion of crossroads  that have to be equipped with relays so as to ensure connectivity. We have also proposed a new cost model for D2D relay deployment, allowing one to more reliably quantify  the necessary investments in relays as a prelude to deploying a relay-assisted D2D network. As a main application, we studied a practical uberization scenario where a neo-operator relies on D2D only to set up its network. We demonstrated that a return on investment with such a strategy  is feasible in a reasonable amount of time, even when the service is funded by low fees. The generic nature of our cost model can help traditional or new actors of the telecommunication sectors to make strategical decisions regarding investments in D2D technology. 

Numerous model extensions are possible. On the geometric point of view, different street systems could be considered, especially those of North American or African cities. On the economic point of view, different scenarios could be studied, e.g. the one of a traditional operator willing to offer D2D as an additional service to its existing customers and hope to get a coverage extension from it. Finally, on the technological point of view, the study of more general shadowing effects, interference or user mobility would lead to more realistic predictions, of greater strategical importance for operators.
\label{s.Conclusions}

\appendix[Computation of the typical crossroad surface ]
 It is  well-known that $\mathscr{S'}(l,\alpha,\beta)=\pi R^2$, where $R$ is the circumcircle radius of the triangle $ABC$ given  by:
\begin{equation*}
    R=\frac{AB \, AC \, BC}{4\mathscr{S}(l,\alpha,\beta)}
\end{equation*}
\noindent Note first that by symmetry arguments in Figure~\ref{fig:streets}, it is easily seen that $(OA)$ is the bissector of the angle spanned by the lines $(d_1)$ and $(d_2)$.  A bit of trigonometry henceforth implies that:
\begin{equation*}
    OA = \frac{l}{2  \sin \left( \frac{\alpha}{2}\right) } \quad ; \quad 
    OB = \frac{l}{2  \sin \left( \frac{\beta}{2}\right) } 
\end{equation*}
And so, by the cross-product formula, the area of the triangle $OAB$ is given by:
\begin{align*}
    S_{OAB} &= \frac{1}{2}OA \, OB \, \sin \widehat{(OA,OB)} \\ &=\frac{l^2}{8\sin\left(\frac{\alpha}{2}\right)\sin\left(\frac{\beta}{2}\right)}\sin\left(\frac{\alpha+\beta}{2}\right).
\end{align*}
In the same way, by a circular permutation of $\alpha,\beta$ and $\delta$, we get analogous formulae for the areas of triangles $OBC$ and $OCA$. Noting that $\delta=2\pi -\alpha -\beta$, we therefore get the surface $\mathscr{S}(l,\alpha,\beta)$ of the triangle $ABC$:
\begin{align}
   \notag \mathscr{S}(l,\alpha,\beta) &= S_{OAB}+S_{OBC}+S_{OCA} \\ \notag  &= \frac{l^2}{4}\left[\cot \frac{\alpha}{2}+ \cot \frac{\beta}{2} + \cot \frac{\delta}{2}\right] \\ \label{crossroad-surface-triangle}  &=  \frac{l^2}{4}\left[\cot \frac{\alpha}{2}+ \cot \frac{\beta}{2} - \cot \frac{\alpha+\beta}{2}\right].
\end{align}


To derive a formula $\mathscr{S}'(l,\alpha,\beta)$ we now only need to compute the side lengths, which is easily done by the law of cosines. For example:
\begin{align*}
    AB^2 &= OA^2 + OB^2 -2 \, OA \, OB \, \cos\left(\widehat{OA,OB}\right) \\ &= \frac{l^2}{4\sin^2 \frac{\alpha}{2}}+\frac{l^2}{4\sin^2 \frac{\beta}{2}}-2\frac{l^2}{4\sin \frac{\alpha}{2}\sin \frac{\beta}{2}}\cos\left(\frac{\alpha+\beta}{2}\right). 
\end{align*}
Factorising by $l^2/4$, expanding the last cosine and using the classical trigonometric identity $1=\cos^2 x + \sin^2 x$ yields:

\begin{equation}
    \label{length-AB}
    AB^2 = \frac{l^2}{4}\left[\left(\cot \frac{\alpha}{2}-\cot \frac{\beta}{2} \right)^2 +4\right].
\end{equation}

Likewise, by circularly permutating $\alpha,\beta$ and $\delta$, and using the fact that $\delta=2\pi - \alpha - \beta$, we get similar formulae for $BC^2$ and $AC^2$. Therefore, 
the area $\mathscr{S}'(l,\alpha,\beta)$ of the circumcircle is given by:


\begin{equation}
\label{crossroad-surface-circumcircle}
 \mathscr{S}'(l,\alpha,\beta)=\pi R^2 = \pi \frac{AB^2 \, BC^2 \, AC^2}{16\mathscr{S}(l,\alpha,\beta)^2},
 \end{equation}
 where $\mathscr{S}(l,\alpha,\beta)$ and $AB$ are respectively given by equations~\eqref{crossroad-surface-triangle} and \eqref{length-AB}. $BC$ and $AC$ are respectively obtained by the circular permutations $(\alpha,\beta) \mapsto (\beta,\delta=2\pi-\alpha-\beta)$ and $(\alpha,\beta) \mapsto (\delta=2\pi-\alpha-\beta,\alpha)$ in~\eqref{length-AB}.

\section*{Acknowledgments}
The authors thank Veluppillai Chandrakumar for fruitful discussions on the cost model.


\bibliographystyle{IEEEtran}
\bibliography{IEEEexample.bib}

\end{document}